\documentclass[twocolumn]{webofc}
\usepackage[varg]{txfonts}   % Web of Conferences font
\begin{document}
\title{High-energy neutrino interaction physics with IceCube}

%\author{\firstname{Spencer} \lastname{Klein}\inst{1,2}\fnsep\thanks{\email{srklein@lbl.gov}}
%\and \firstname{for the}\lastname{IceCube Collaboration}
        % etc.
%}

\author{Spencer Klein\inst{1,2}\fnsep\thanks{\email{srklein@lbl.gov}} for the IceCube Collaboration\thanks{\protect\url{http://icecube.wisc.edu/collaboration/authors/current}}}

\institute{Lawrence Berkeley National Laboratory, Berkeley CA 94720 USA \and
University of California, Berkeley, CA, 94720 USA
}

\abstract{
Although they are best known for studying astrophysical neutrinos, neutrino telescopes like IceCube can study neutrino interactions, at energies far above those that are accessible at accelerators.   In this writeup, I present two IceCube analyses of neutrino interactions at energies far above 1 TeV.   The first measures neutrino absorption in the Earth, and, from that determines the neutrino-nucleon cross-section at energies between 6.3 and 980 TeV.   We find that the cross-sections is 1.30 $^{+0.21}_{-0.19}$ (stat.) $^{+0.39}_{-0.43}$ (syst.) times the Standard Model cross-section.  We also present  a measurement of neutrino inelasticity, using $\nu_\mu$ charged-current interactions that occur within IceCube.   We have measured the average inelasticity at energies from 1 TeV to above 100 TeV, and found that it is in agreement with the Standard Model expectations.  We have also performed a series of fits to this track sample and a matching cascade sample, to probe aspects of the astrophysical neutrino flux, particularly the flavor ratio.  
 }
\maketitle
\section{Introduction}
The IceCube observatory has observed neutrinos with energies well above 2 PeV, far beyond the 500 GeV reached by the most energetic terrestrial neutrino beams.  These neutrinos have given us significant insight into the high-energy universe, particularly regarding astrophysical accelerators.  However, these same neutrinos also
offer us the opportunity to extend neutrino interaction studies to much higher energies.   

IceCube consists of 86 vertical strings of optical sensors (digital optical modules, or DOMs) which were lowered into holes drilled into the Antaractic ice cap at the South Pole \cite{Halzen:2010yj,Aartsen:2016nxy}.  The array covers a surface area of 1 km$^2$.   Each string is instrumented with 60 DOMs.  On 78 strings, the DOMs are positioned every 17 m, between a depth of 1450 and 2450 m.   On the remaining strings, the DOMs are deployed with a 7 m spacing near the bottom of the array. 

Each DOM \cite{Abbasi:2008aa} consists of a 25 cm photomultiplier tube and data acquisition (DAQ) electronics, in a clear glass pressure vessel.   Two waveform digitizer systems record the arrival times of most photoelectrons.  A calibration system maintains the timing calibration for all of the DOMs to within 3 nsec  \cite{Achterberg:2006md}.     All of the data is sent to the surface, where software triggers find groups of time-correlated hits, and send the data to a processor farm for on-line reconstruction. 

One of the major challenges in studying neutrino interactions with IceCube is understanding the beam, and accounting for its uncertainties.  The beam has three components, conventional and prompt atmospheric neutrinos, and astrophysical neutrinos, each with different, often energy-dependent flavor composition and neutrino:antineutrino ratios.  Here, we focus on $\nu_\mu$, so are not limited by the uncertainties in flavor composition.   All three components have somewhat different $\nu:\overline\nu$ ratios.  Since $\nu$ and $\overline\nu$ are indistinguishable in neutrino telescopes, but interact with different cross-sections and produce different inelasticity distributions, the $\nu:\overline\nu$ ratio is a significant systematic uncertainty.

\section{Cross-section measurement}

In the Standard Model neutrinos interact via charged-current (CC) and neutral-current (NC) deep inelastic scattering.  The cross-sections for these process increase with increasing energy.    As Volkova and Zatsepin \cite{Volkova}  first pointed out, at energies above about 40 TeV, the Earth becomes opaque to neutrinos, so one can use energetic neutrinos to probe the interior of the Earth.   Or, one can turn this around, assume that the Earth density profile is known, and use absorption to measure the neutrino-nucleon cross-section.  For long chords through the Earth, as are studied here, the standard ``Preliminary Earth Reference Model" \cite{PREM} contributes less than a 1\% systematic uncertainty to the cross-section measurement. 

\begin{figure}[h]
\centering
\includegraphics[width=8.5 cm,clip]{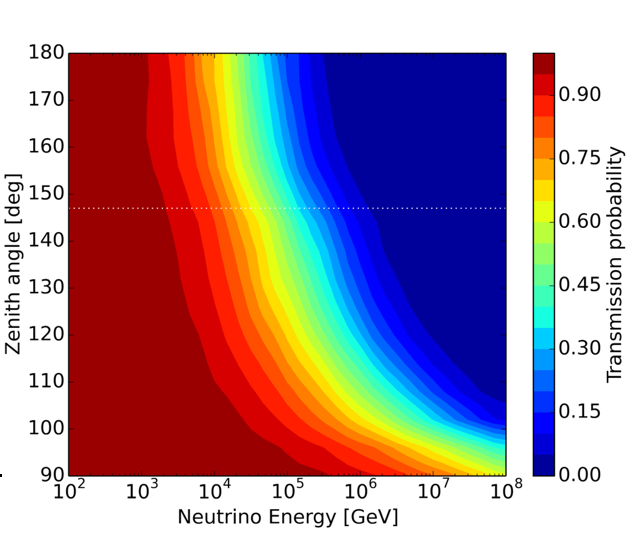}
\caption{The $\nu_\mu$ transmission probability as a function of neutrino energy and zenith angle.  The horizontal dashed white line shows the core mantle boundary.  The high core density produces a noticeable inflection at that point. 
From Ref. \cite{Aartsen:2017kpd}.
}
\label{fig:transmit}       % Give a unique label
\end{figure}

Figure \ref{fig:transmit} shows the transmission probability, as a function of zenith angle and neutrino energy.   Absorption manifests itself as a change in the zenith angle distribution with increasing neutrino energy.   Beyond Standard Model (BSM) processes might also contribute to the  cross-section, further increasing the rise with energy.   Two phenomena that can increase the cross-section are models that involve leptoquarks \cite{Romero:2009vu} and those with additional, tightly-rolled-up spatial dimensions \cite{AlvarezMuniz:2002ga}.   
 
We measured the cross-section using 1 year of data from IC-79, a sample of 10,784 upward-going $\nu_\mu$ events \cite{Aartsen:2015rwa} with measured muon energy above 1 TeV \cite{Abbasi:2012wht}.    These events were then binned in two dimensions, cos($\theta_z$) and muon energy.  The resulting histogram was then fit to a model that included the three fluxes from conventional and prompt atmospheric neutrino and astrophysical neutrinos. 

The cross-section was determined by a maximum-likelihood fit which had the cross-section as a free parameter  \cite{Aartsen:2017kpd,Sandydissertation}.  It was assumed to be a multiple of the Standard Model cross-section $\sigma_\nu$, with $R=\sigma_{\nu}/\sigma_{\rm CSMS}$, where $\sigma_{\rm CSMS}$ is the Standard Model cross-section, as computed in \cite{CooperSarkar:2011pa}.  This calculation is done in next-to-leading order perturbative QCD, with DGLAP evolution used to extrapolate the parton distribution functions to the low-$x$ region.   A similar calculation, by Connolly, Thorne and Waters, found similar results \cite{Connolly:2011vc}.

 In the analysis, the charged-current (CC) and neutral-current (CC) cross-sections are assumed to vary in parallel \cite{Klein:2013xoa}.   In NC interactions, the neutrinos lose energy, but are not absorbed.  The fit accounted for these interactions, by treating propagation through the Earth as a two-dimensional problem, with one dimension for the entering neutrino energy, and the other for its energy when it reaches IceCube.   The propagation was calculated for different cross-sections, as a function of these energies and zenith angle, and the fitter interpolated between the nearest cross-sections.  The absorption calculation neglected nuclear shadowing, which can reduce the per-nucleon cross-section of heavy nuclei \cite{Armesto:2006ph}.  It also neglected electromagnetic interactions, whereby the neutrino fluctuates to a $\mu$ and a $W^\pm$, with the $\mu$ then interacting with the Coulomb field of the target nucleus.   Overall, these should both be less than 10\% effects. 

The fit also included seven nuisance parameters to account for uncertainties in the neutrino fluxes, plus one for the DOM efficiency.  These were the normalizations for the conventional and prompt atmospheric fluxes and the astrophysical flux, the cosmic-ray spectral index, and the $K/\pi$ and $\nu/\overline\nu$ ratio for conventional neutrinos, and the astrophysical flux spectral index.   An additional nuisance parameter accounts for uncertainties in the overall DOM optical sensitivity.

The fit found a cross-section multiplier of $R=1.30^{+0.30}_{-0.26}$.  This is the statistical uncertainty, plus some systematic uncertainty due to the nuisance parameters.  We isolated the statistical uncertainty by fixing the nuisance parameters to their preferred values and rerunning the fit.   We then determined the systematic uncertainties associated with the fit by subtracting, in quadrature, the statistical uncertainty from the total fit uncertainty.   The total systematic uncertainty includes some contributions from factors which were not included in the fit: uncertainties about the optical properties of the ice ($^{+0.30}_{-0.38}$), uncertainties in the density distribution of the Earth ($\pm 0.01$), latitude-dependent variations in production rate due to temperature ($^{+0.00}_{-0.04}$),  uncertainties in the angular acceptance of the IceCube DOMs ($^{+0.04}_{-0.00}$) and finally uncertainties in the spectral indices of the prompt and astrophysical spectral indices.  The latter were already included in the fit; this additional uncertainty was included to account for the tension between the spectral indices observed by contained event studies \cite{Aartsen:2017mau} and those from through-going muons \cite{Aartsen:2016xlq}.  We then found the total systematic uncertainty by adding these factors, in quadrature to the systematic error associated with the fit.  This led to the final result, that the cross-sections is 1.30 $^{+0.21}_{-0.19}$ (stat.) $^{+0.39}_{-0.43}$ (syst.) times the Standard Model cross-section. 

\begin{figure}[h]
\centering
\includegraphics[width=8.5 cm,clip]{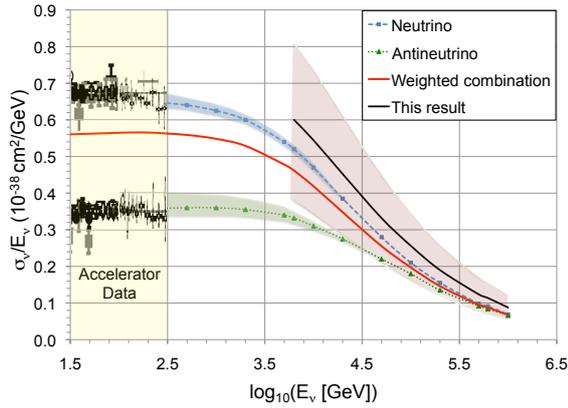}
\caption{The IceCube neutrino cross-section (divided by neutrino energy) measurement (black line and shaded band), along with previous accelerator measurements (points with error bars).  The blue line (with squares) and the green line (with triangle symbols) show the predictions \cite{CooperSarkar:2011pa} 
for $\nu_\mu$ and $\overline\nu_\mu$ respectively, while the red line is the expectation for the mixture seen in IceCube.  The cross-sections scale linearly with energy up to about 1 TeV in energy, where the finite $W^\pm$ and $Z^0$ masses moderate the rise with increasing energy.  From Ref. \cite{Aartsen:2017kpd}.
}
\label{fig:cross-section}       % Give a unique label
\end{figure}

We determined the energy range for which this measurement applied by studying the change in likelihood as we turned off Earth absorption, first starting from very low energies, working upward, and then starting from very high energies, working downward.  The points where the likelihood worsened by $\-2\Delta LLH =1$ gave us the minimum and maximum energies respectively, 6.3 TeV and 980 TeV.  Figure \ref{fig:cross-section}  shows this result, along with previous lower-energy results from accelerator experiments.  The data does not show a large rise, as would be expected in some BSM theories, particularly those involving leptoquarks or additional rolled-up spatial dimensions.

Similar analyses have been done using contained cascade events \cite{Bustamante:2017xuy,yiqian}.  These analyses suffer from much more limited statistics, so the statistical errors are much larger.  On the other hand, the cascade energies are much better known, so it is easier to measure the cross-section in multiple energy bins. 

\section{Inelasticity measurement}

If BSM processes contribute to the cross-section, there is no reason to expect the form of the interactions to be similar to those from Standard Model processes, so they are likely to look different in a detector.  One measure of neutrino interactions is the distribution of inelasticity, $y$, the fraction of the energy that is transferred from an incident neutrino into a struck nucleon; the remainder of the energy is transferred to the created muon.   The inelasticity distribution is well predicted in the Standard Model.   The average $y$ decreases from about 0.5 at low energies (100 GeV) to about 0.25 for 100 PeV neutrinos.   Beyond tests of the Standard Model, the inelasticity has a number of other applications.  Since some $\nu_\tau$ interactions will appear as starting tracks, it is sensitive to the flavor ratio for astrophysical neutrinos.  It is also sensitive to the $\nu:\overline\nu$ ratio, since $\nu$ and $\overline\nu$ have somewhat different inelasticity distributions, albeit only for neutrino energies below about 10 TeV.  

We measured the inelasticity distribution of  CC $\nu_\mu$ interactions, using a sample of contained events \cite{newpaper,Gary} found in 5 years of IceCube data.   We select a sample of contained events using an outer layer veto, following Ref. \cite{Aartsen:2015ivb}, but with some changes to reduce the energy threshold \cite{Aartsen:2014muf}.  A boosted decision tree (BDT) is then used to further reduce background, based on 15 relevant variables, including direction (when reconstructed as a track) position, energy and length; variables based on reconstruction as a cascade are also used.   The BDT had separate output for events classified as starting tracks and as cascades. The cascades provide an important comparison point for some of the fits presented  below.   

\begin{figure}[h]
\centering
\includegraphics[width=\linewidth]{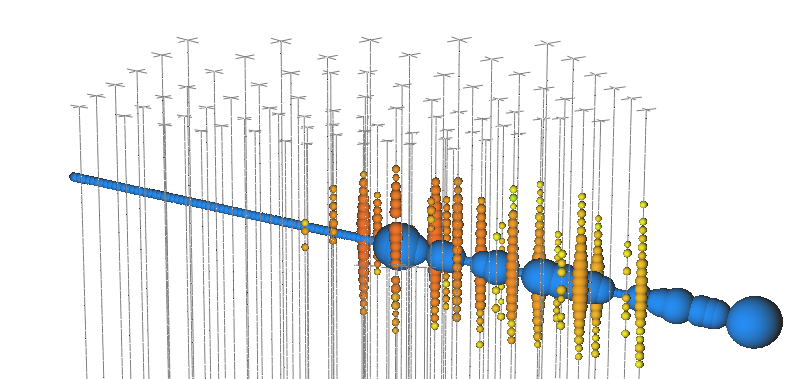}
\vskip .2 in
\includegraphics[width=\linewidth]{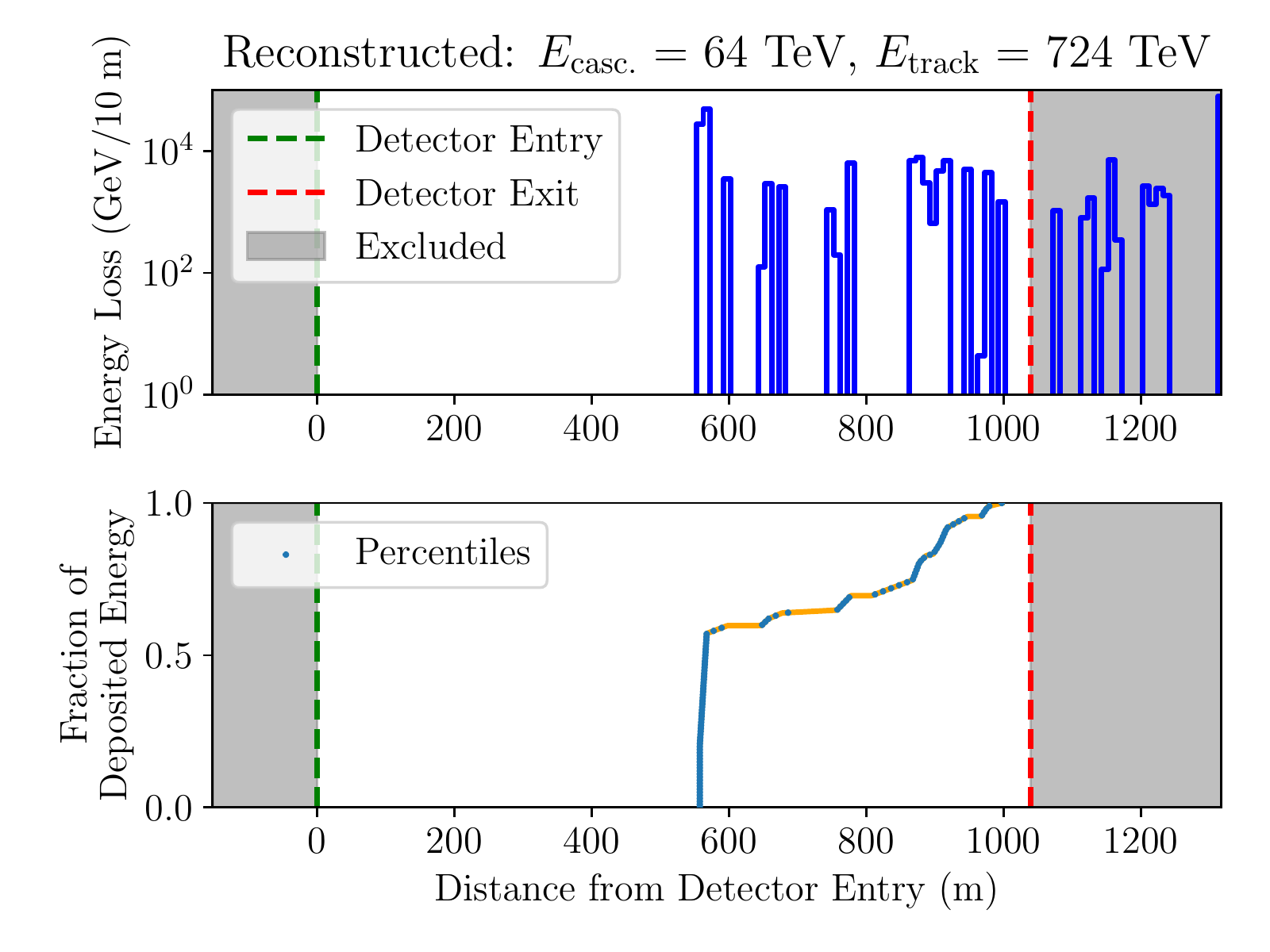}
\caption{The highest energy event found in the starting track analysis.  Top: Each colored sphere shows a DOM, with the size of the sphere showing the number of photons observed by the DOM.  The blue line shows the reconstructed track, which starts in the middle of the detector; the blue spheres along the track show the reconstructed energy deposition.  The middle plot shows this energy deposition, while the bottom panel shows the integrated energy deposition, after energy loss outside the detector is removed; this is input to the random forest that determines the cascade and track energies.   The cascade and muon energies are estimated to be 64 and 724 TeV respectively, for a total neutrino energy of 788 TeV.  Because of the small cascade energy, the visible inelasticity is small, $y_{\rm vis.}=0.08$.  
From Ref. \cite{newpaper}.}
\label{fig:event}       % Give a unique label
\end{figure}

In 5 years of IceCube data, the analysis selected 2650 starting tracks and 965 cascades.  Figure \ref{fig:event} shows the most energetic starting track.  

The cascade events were reconstructed using a standard IceCube maximum likelihood fitter \cite{Aartsen:2015ivb}.  For the starting tracks, the track directions were reconstructed in the usual manner, but the cascade and track component energies were estimated using machine learning techniques.    A random forest was used to separately estimate the starting track cascade energy $E_{\rm casc.}$ and muon energy $E_\mu$, based on the integrated energy loss profile of the track, as shown in Fig. \ref{fig:event}.    The total visible energy is $E_{\rm vis.}= E_{\rm casc.} + E_\mu$, and the visible inelasticity is $y_{\rm vis.} = E_{\rm casc.}/E_{\rm vis.}$.  The label 'visible' is to account for energy loss to neutrinos and other processes that are invisible in a Cherenkov detector.   Also, the event selection and classification lose efficiency when $y$ is near 0. or 1.   When $y\approx 0$ (very little cascade energy), the total energy deposited in the detector is lower, and the event is less likely to be observed, particularly for lower energy neutrinos.   When $y\approx 1$ (very little track energy), then the event may be mis-classified as  a cascade.    Overall, the RMS resolution  is 0.18 for $\log_{10} E_{\rm vis.}$ and 0.19 for $y_{\rm vis.}$. 

\begin{figure*}[h]
\centering
\includegraphics[width=0.9\linewidth]{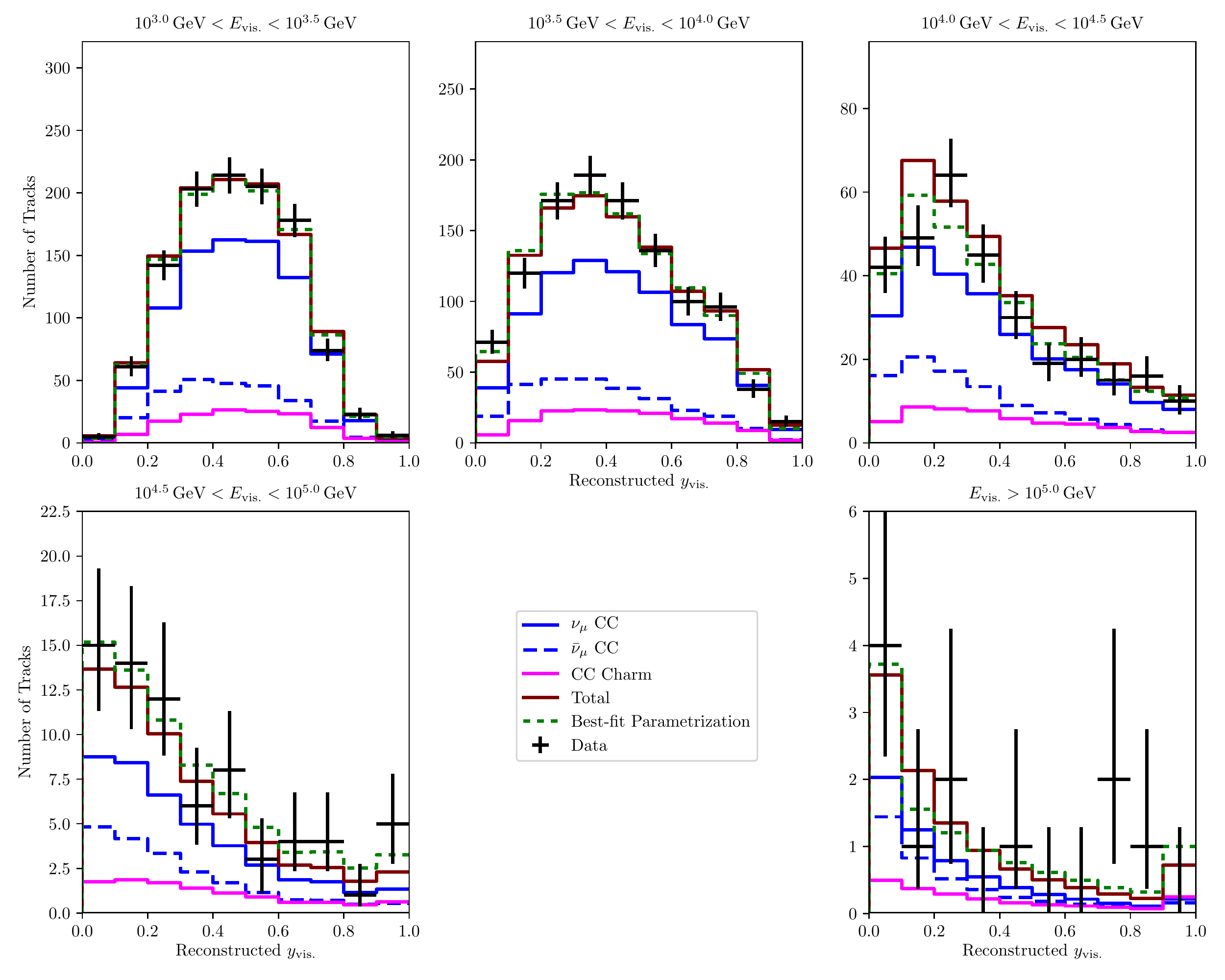}
%\vspace*{5cm} 
\caption{The reconstructed visible inelasticity distribution in five different bins of reconstructed energy.  Observed data are shown in black, and the result of fitting the distribution to the parameterization of Eq.~\ref{eq:nu_cc_xs_y_param} and the best-fit neutrino flux parameters is shown with dashed green lines.  The prediction of the CSMS differential CC cross section \cite{CooperSarkar:2011pa} are shown for neutrinos with solid blue lines and antineutrinos with dashed blue lines.  The total CC charm contribution is shown in magenta, illustrating its flatter inelasticity distribution. From Ref. \cite{newpaper}.}
\label{fig:inel_dists}
\end{figure*}

Figure \ref{fig:inel_dists} shows the visible inelasticity distribution, in 5 energy bins.   Also shown is a fit, based on a conventional, prompt and astrophysical neutrino flux mixture that is similar to that used for the cross-section measurement.   The inelasticity distributions used for the fit are from Ref. \cite{CooperSarkar:2011pa}.  

\begin{figure}
  \centering
  \includegraphics[width=\linewidth]{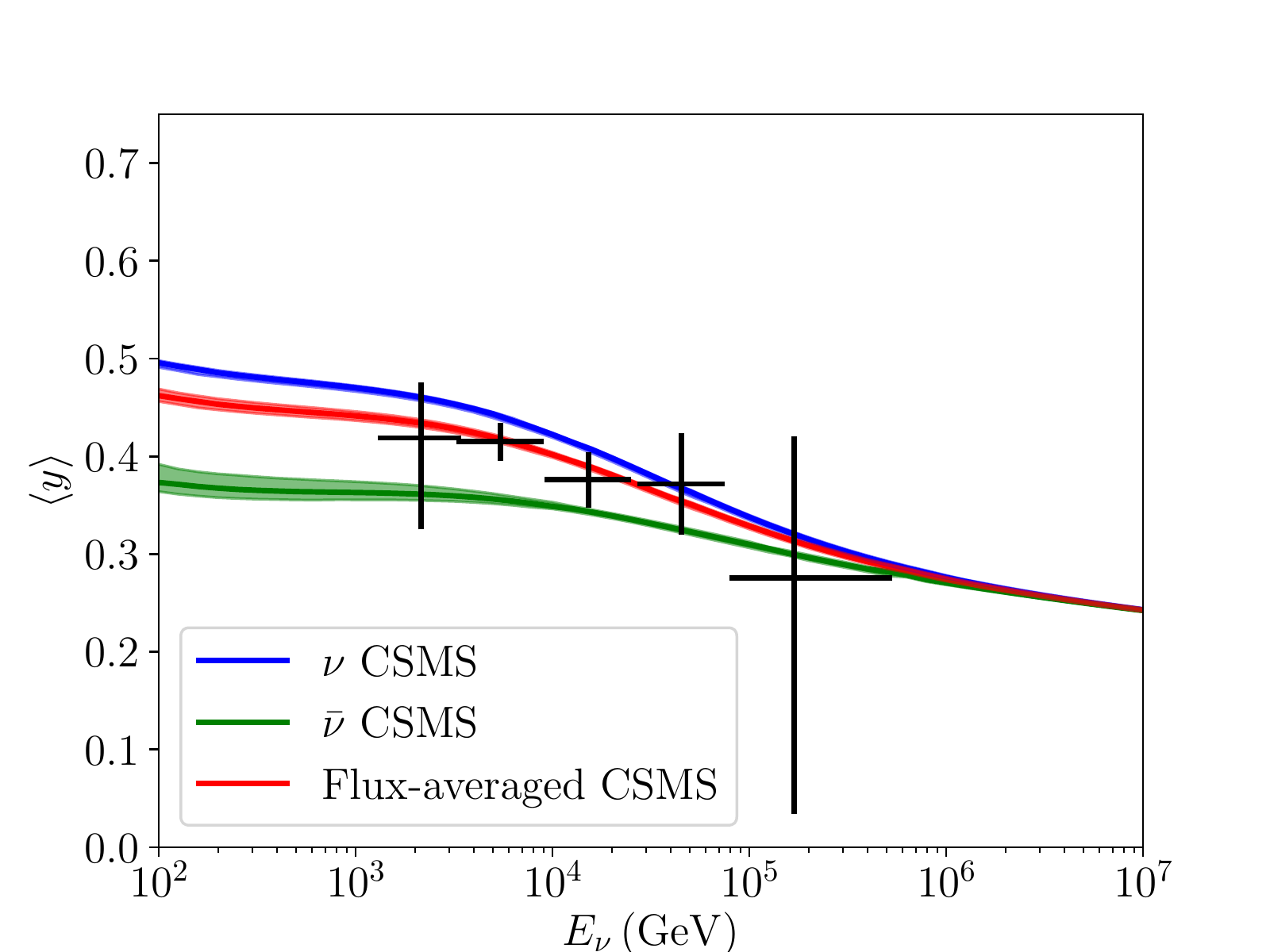}
  \caption{The mean inelasticity in five energy bins.  The vertical error bars show the $68\%$ confidence interval for the mean inelasticity, and horizontal error bars cover the expected central $68\%$ of neutrino energies in each bin.  The Standard Model prediction \cite{CooperSarkar:2011pa} is shown in blue for neutrinos and in green for antineutrinos.  The red lines show the flux-averaged mean inelasticity, based on a cosmic-ray model which extrapolates the HKKMS calculation \cite{Honda:2006qj} upward in energy, accounting for the knee.  From Ref. \cite{newpaper}.}
\label{fig:split_fit_inel}
\end{figure}

Because of the uneven efficiency, we do not present unfolded $y$ distributions.  Instead, we parameterize $d\sigma/dy$ in terms of two variables, $\epsilon$ and $y$:
\begin{equation}
\frac{d\sigma}{dy} = N [1+\epsilon(1-y)^2)y^{\lambda-1}]
\label{eq:nu_cc_xs_y_param}
\end{equation}
where $N$ is the normalization.  The parameterization is motivated by the expected behavior of sea-quark parton distributions at low Bjorken$-x$: $xq(x,Q^2)\propto A(Q^2)x^{-\lambda}$.   Unfortunately, $\epsilon$ and $\lambda$ are highly correlated, so we present our results in terms of $<y>$ and $\lambda$, which are much less correlated.  Figure \ref{fig:split_fit_inel} shows $<y>$ in the five energy bins with the Standard Model predictions; the agreement is good. 

\subsubsection{Other physics from inelasticity}

The addition of inelasticity to starting event analyses allows us to probe some additional areas of neutrino interactions, and also to better constrain the astrophysical neutrino flux.

\begin{figure}
  \centering
  \includegraphics[width=\linewidth]{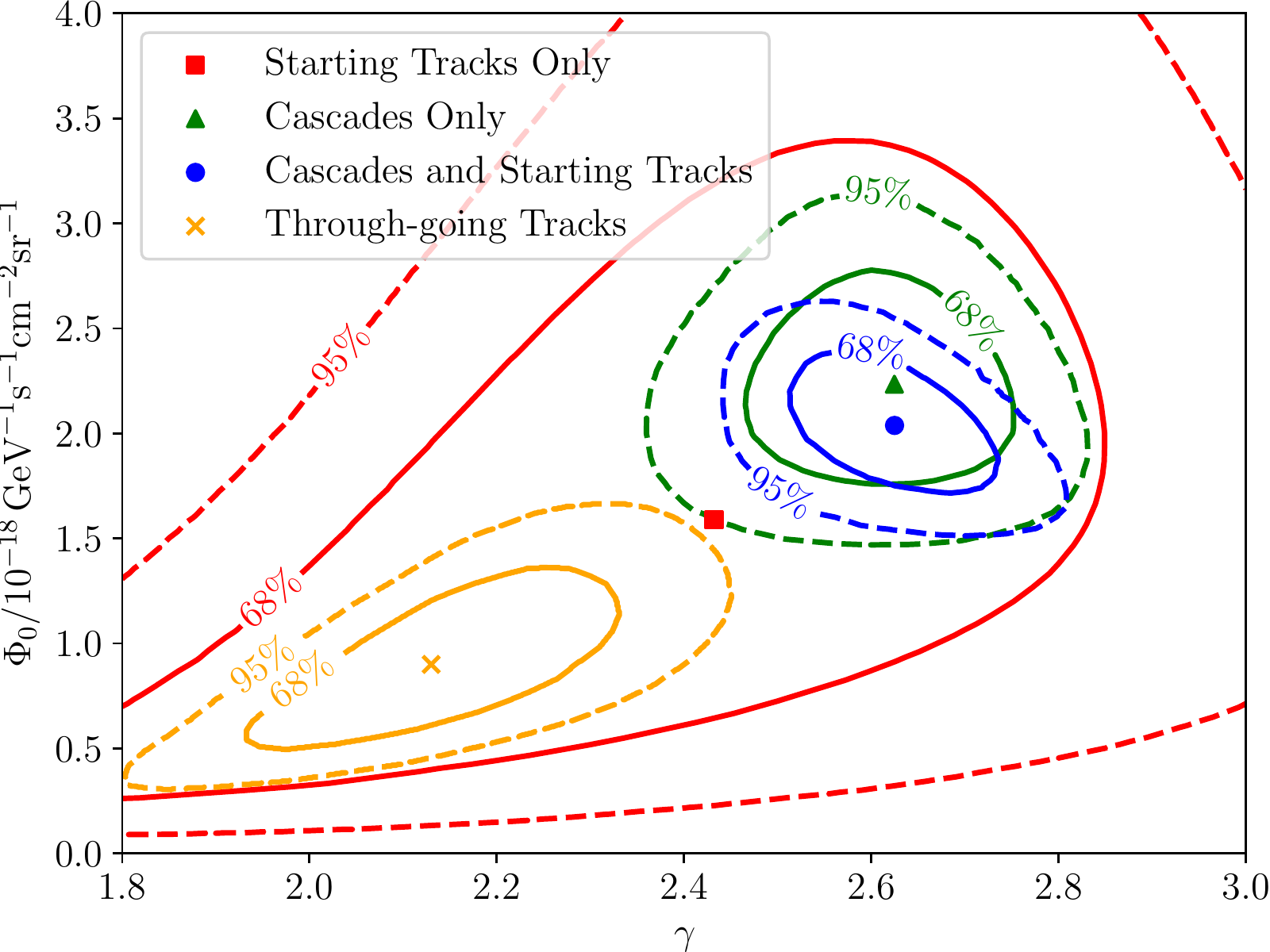}
  \caption{Contours for the confidence region for astrophysical power law index (x axis) and flux normalization (y-axis) for the current starting-track (red squares) and cascades (green triangles), and combined fit (blue circles), along with the previous through-going track study (orange x) \cite{Aartsen:2017mau}.  From Ref. \cite{newpaper}.
  }
\label{fig:contour}
\end{figure}

The fit described above found an astrophysical spectral index of $\gamma=2.62\pm 0.07$, in good agreement with previous IceCube contained event and cascade analyses \cite{Aartsen:2017mau} , but softer than recent measurements using through-going muons  \cite{Aartsen:2016xlq}.     The high quality of the starting track reconstruction allows us to perform another fit, where the astrophysical fluxes (including indexes) were allowed to float separately for the starting tracks and cascades. When this is done, we find $\gamma=2.43^{+0.28}_{-0.30}$ for starting tracks, and $\gamma=2.62\pm 0.08$ for the cascades in the sample.  The combined-sample $\gamma$ is driven by the cascades, and the starting-track sample has a much larger uncertainty.   The starting-track result is mid-way between the previous starting event samples \cite{Aartsen:2017mau} and the previous through-going muon samples \cite{Aartsen:2016xlq}.   Figure \ref{fig:contour} shows the contour plots for likelihoods for these spectral indices.

\begin{figure}
  \centering
  \includegraphics[width=\linewidth]{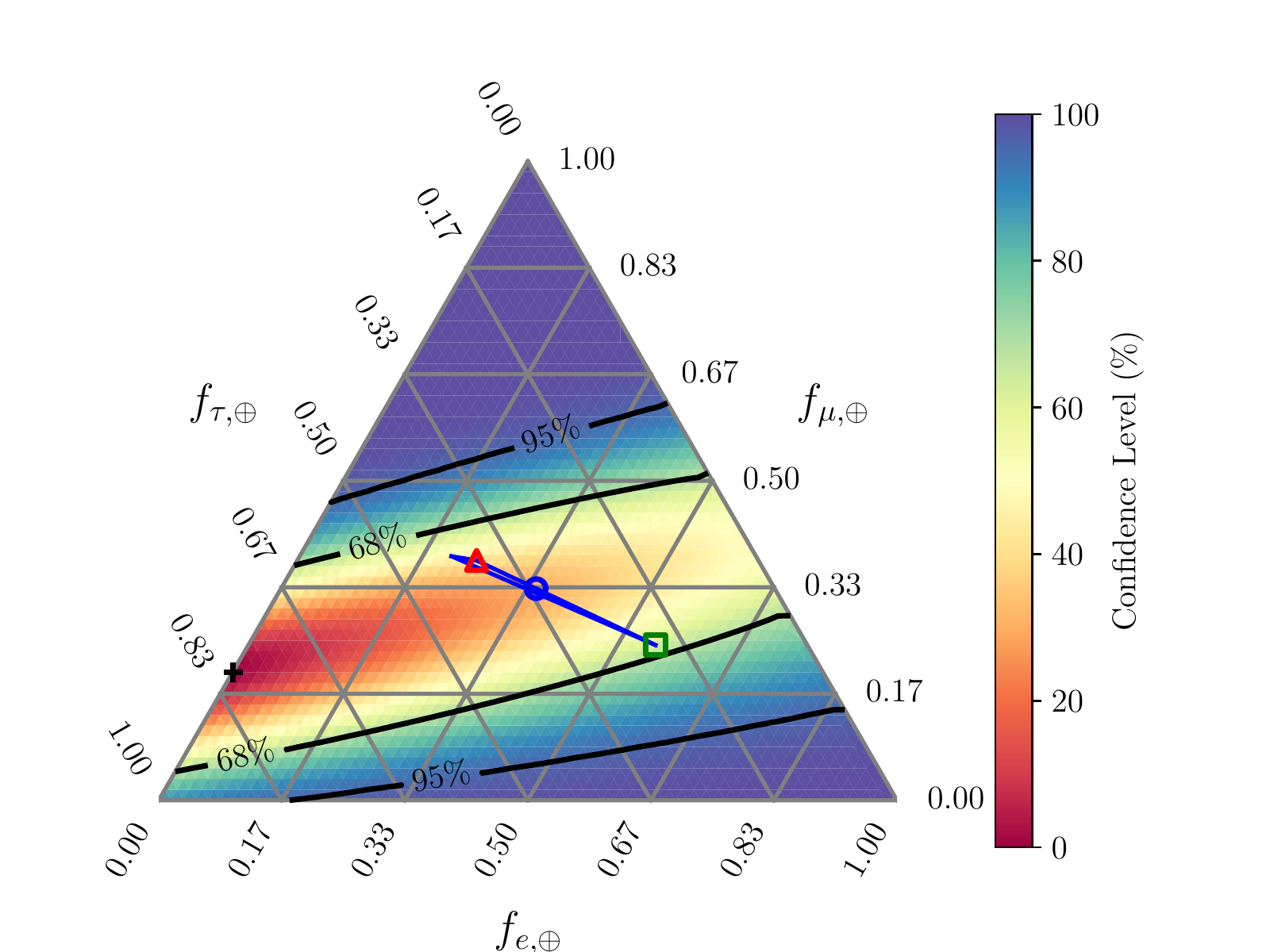}
  \caption{Confidence levels for various astrophysical neutrino flavor ratios, as observed on Earth.  The labels on the three sides of the triangle correspond to the tilted lines along the interior.   Each point in the interior corresponds to a given ($\nu_\tau$, $\nu_\mu$,$\nu_e$) ratio; the fractions can be determined by following the lines to the upper left, side and downward to the left respectively.  The best fit is 79\% $\nu_\tau$ and 21\% $\nu_e$, while the colors show the likelihood contours.    The red triangle, blue circle and green square, and their connecting lines, show the expectations for conventional acceleration mechanisms, assuming Standard Model oscillations in-transit.
 From Ref. \cite{newpaper}.}
\label{fig:triangle}
\end{figure}

Inelasticity also has some discriminating power for $\nu_\tau$.   $\nu_\tau$ decaying to muons are classified as starting tracks, but with a lower visible energy and a higher visible inelasticity distribution then $\nu_\mu$ CC interactions.   So, inelasticity helps to constrain the astrophysical neutrino flavor triangle.  Figure \ref{fig:triangle} shows the flavor triangle derived here, using both the cascades and starting tracks.   Two corners of the triangle, 100\% $\nu_\mu$ and 100\% $\nu_e$ are ruled out at more than 5$\sigma$ confidence level.    Unfortunately, the analysis does not yet have the sensitivity to distinguish between different conventional acceleration scenarios. 

The allowed region is tighter than in previous studies of contained events \cite{Aartsen:2015ivb}.   The allowed region is broader than was found by a previous combined fit \cite{Aartsen:2015knd}, but that fit used cascade and track samples with rather different energies, and the result necessarily assumes that the astrophysical flux is a single power law.   Here, the two samples have similar energies, with 90\% of the starting track events having energies between 740 GeV and 45 TeV, while 90\% of the cascade events fall in the range from 1.1 TeV and 53 TeV.  So, this result is largely independent of the astrophysical energy spectrum.

At energies below about 10 TeV, $\nu_\mu$ and $\overline\nu_\mu$ have somewhat different inelasticity distributions, because the $W^\pm$ and $Z^0$ bosons couple to quarks and antiquarks in the target respectively.   This is energy range is too low to probe the astrophysical flux, but it is sensitive to atmospheric neutrinos.   From a fit where the conventional atmospheric $\nu_\mu:\overline\nu_\mu$ is allowed to vary, we find that the $\nu_\mu:\overline\nu_\mu$ ratio is $0.77^{+0.44}_{-0.25}$ times the fraction in the extrapolated HKKMS calculation \cite{Honda:2006qj}.  The sensitive range for this fit is 770 GeV up to 21 TeV.  

Another fit explored charm production in neutrino events.  Charm quarks are produced when neutrinos interact with a strange quark in the target nucleus.  Because there are no strange valence quarks, they have a different parton distribution function from up and down quarks.  This leads to a flatter inelasticity distribution.   From a fit where the charm production fraction was allowed to float, we find that that charm production is $0.93^{+0.73}_{-0.59}$ times the Standard Model CSMS prediction.  Charm production is observed at more than a 90\% confidence level.  This measurement is for neutrino energies from 1.5 to 340 TeV (90\% coverage range).

\section{Conclusions}

IceCube has made two measurements of neutrino interaction physics.  We have observed neutrino absorption in the Earth, and, from that, determined the cross-section at energies between 6.3 and 980 TeV.  We have isolated a sample of starting tracks and measured their energies.  From that, we have made measurements of the inelasticity distributions and mean inelasticity.  We have performed an additional series of fits to study the astrophysical neutrino spectrum and flavor composition, determined the neutrino:antineutrino ratio of atmospheric neutrinos, and observe charm production in neutrino events.

Looking ahead, we expect to use more data, to make a more precise cross-section measurement, divided into neutrino energy bins, and thereby extend the measurement to higher energies.  These data can also be used to more precisely constrain BSM physics. 

The inelasticity measurements can be combined with other variables to better constrain the astrophysical neutrinos spectrum and composition, in  a single fit.  The proposed IceCube-Gen2 extension \cite{Ackermann:2017pja} and/or KM3NeT 2.0 detector \cite{Adrian-Martinez:2016fdl} would have significantly increased data collection capabilities, and could extend these measurements to still higher energies.   Looking further ahead, instruments that detect the radio waves from neutrino-induced showers will have much larger effective volumes, in the 100~km$^3$ range \cite{Allison:2015eky,Barwick:2014pca}.  These detectors can collect a sample of neutrinos with energies above $10^{19}$ eV, and thereby study the neutrino cross-section at energies above those accessible at the Large Hadron Collider. 

This work was supported in part by U.S. National Science Foundation under grants PHY-1307472 and the U.S. Department of Energy under contract number DE-AC02-05-CH11231.

\end{document}